\documentclass[amssymb,prb,twocolumn,showpacs,superscriptaddress]{revtex4}
\usepackage{graphicx}
\usepackage{bm}
\usepackage{amsmath}
\usepackage[latin1]{inputenc}

\begin{document}

\title{Finite-size effects on the dynamic susceptibility of CoPhOMe\\
single-chain molecular magnets in presence of a static magnetic field}


\author{M. G. Pini}
\affiliation{Istituto dei Sistemi Complessi, Consiglio Nazionale
delle Ricerche, I-50019 Sesto Fiorentino, Firenze, Italy }
\author{A. Rettori} \affiliation{Dipartimento di Fisica and CNISM, Universit\`a di Firenze, I-50019
Sesto Fiorentino, Firenze, Italy}
\author{L. Bogani}\affiliation{1. Physikalisches Institut, Universit\"at Stuttgart, D-70550 Stuttgart, Germany}
\author{A. Lascialfari}\affiliation{Dipartimento di Fisica "A. Volta",
Universit\`a di Pavia, I-27100 Pavia,
Italy}\affiliation{Dipartimento di Scienze Molecolari Applicate ai
Biosistemi, Universit\`a di Milano, I-20134 Milano, Italy}
\author{M. Mariani}\affiliation{Dipartimento di Fisica "A. Volta",
Universit\`a di Pavia, I-27100 Pavia, Italy}
\author{A. Caneschi}\affiliation{Dipartimento di Chimica and INSTM, Universit\`a di Firenze,
I-50019 Sesto Fiorentino, Firenze, Italy}
\author{R. Sessoli}\affiliation{Dipartimento di Chimica and INSTM,
Universit\`a di Firenze, I-50019 Sesto Fiorentino, Firenze, Italy}


\date{\today}

\begin{abstract}

The static and dynamic properties of the single-chain molecular
magnet [Co(hfac)$_2$NITPhOMe] are investigated in the framework
of the Ising model with Glauber dynamics, in order to take into
account both the effect of an applied magnetic field and a finite
size of the chains. For static fields of moderate intensity and
short chain lengths, the approximation of a mono-exponential decay
of the magnetization fluctuations is found to be valid at low
temperatures; for strong fields and long chains, a
multi-exponential decay should rather be assumed. The effect of an
oscillating magnetic field, with intensity much smaller than that
of the static one, is included in the theory in order to
obtain the dynamic susceptibility $\chi(\omega)$.
We find that, for an open chain with $N$ spins, $\chi(\omega)$ can
be written as a weighted sum of $N$ frequency contributions, with
a sum rule relating the frequency weights to the static susceptibility
of the chain. Very good agreement is found between the theoretical
dynamic susceptibility and the ac susceptibility measured in moderate
static fields ($H_{\rm dc}\le 2$ kOe), where the approximation
of a single dominating frequency turns out to be valid. For static fields
in this range, new data for the relaxation time, $\tau$ versus $H_{\rm dc}$,
of the magnetization of CoPhOMe at low temperature are also well
reproduced by theory, provided that finite-size effects are included.
\end{abstract}

\pacs{75.50.Xx,75.40.Gb,75.10.Hk,76.20.+q}





\maketitle

\section{Introduction}
\label{Introduction}

It is commonly admitted that the plain Ising Hamiltonian does not
contain any dynamics.\cite{Villain_IsingAF} In fact, when
considering a system of Ising spins, $\sigma_i$, localized at the
sites, $i$, of a lattice
\begin{equation}
\label{IsingFM_general} \mathcal{H}
=-J_I\sum_{ij} \sigma_{i}\sigma_{j},~~~~~~~~~\sigma_i=\pm 1
\end{equation}
\noindent the physically interesting quantities, $\sigma_i$'s,
commute with $\mathcal H$. However, for a system in contact with a
heat bath, a stochastic dynamics can be introduced by means of a
master equation which assumes Markovian processes inducing random
flips between different states. One of the few cases in which the
problem can be solved analytically is for a one-dimensional
lattice, zero external magnetic field, and an opportune choice of
the transition probability, as devised by
Glauber\cite{Glauber_1963} some decades ago.  He calculated the
dynamic susceptibility within a linear response framework, and
found that the uniform magnetization decays exponentially, with a
relaxation time given by an Arrhenius law
\begin{equation}
\label{tau4J}
\tau(T)=\tau_0~ e^{{{4J_I}\over {k_BT}}},
\end{equation}
\noindent where $1/\tau_0$ is the relaxation rate for an isolated
spin. Considering that, at low temperatures, the correlation
length for model (\ref{IsingFM_general}) in one dimension
is given by $\xi\propto e^{{{2J_I}\over {k_BT}}}$,\cite{Baxter}
the dynamic critical exponent results in $z=2.\cite{Cordery_1981}$
Glauber's dynamics has been ever since applied in very different areas,
comprising structural phase transitions\cite{Khoshe}, neural
networks\cite{Bialek}, chemical reactivity,\cite{Skinner} and even
bio-socio-econo-physics.\cite{Stauffer1,Stauffer2}

The recent discovery of single-chain magnets
(SCMs)\cite{Caneschi_2001} spurred renewed interest for Glauber's
dynamics in magnetic nanomaterials. Such systems show magnetic
hysteresis without the onset of three-dimensional magnetic
ordering. At very low temperatures, the relaxation of the
magnetization is so slow that also other very interesting
dynamic phenomena have been observed with unprecedented clarity,
including collective reversal\cite{Vindigni_2005} and quantum
tunneling\cite{QT}. As a result, after the discovery of the
archetypal SCM [Co(hfac)2NITPhOMe] (hereafter denoted
CoPhOMe),\cite{Caneschi_2001} the number of SCM compounds has been
rapidly increasing.\cite{Structure_2006,Rassegna_SCM,
Lapo1,Lapo2,Lapo3,Lapo4,Lapo5,Lapo6}

The strong exchange interaction and one-dimensional character of
CoPhOMe\cite{Caneschi_2001} make it the ideal system where to
observe the long-predicted\cite{Glauber_1963} slow relaxation of
the magnetization. In CoPhOMe, owing to the very high value of the
exchange constant ($J_I/k_B=80$ K), the correlation length $\xi$
is huge, at low temperatures, in a zero magnetic field.
Consequently, the unavoidable presence of even a small density of
defects causes the chain to break into finite segments whose
average length, $\bar{L}$, can be much smaller than the
correlation length, $\xi$. In such a finite-size regime, $\bar{L}
\ll \xi$, the dynamics of the system in zero field is
modified\cite{DaSilva,Luscombe} with respect to Glauber's analysis
of the infinite chain.\cite{Glauber_1963} In particular the
relaxation time, measured in CoPhOMe using ac susceptibility and
SQUID magnetization decay techniques, was found to follow an
Arrhenius law with a halved energy
barrier\cite{Caneschi_2002,PRL_2004,Vindigni_2004}
\begin{equation}
\label{tau2J}
\tau(T)=\tau_0(\bar{L})~e^{{{2J_I}\over {k_BT}}},
\end{equation}
in agreement with theoretical predictions of finite-size
effects.\cite{DaSilva,Luscombe}

These effects were systematically investigated in CoPhOMe by
introducing non-magnetic impurities.\cite{Bogani_2005} For
nominally pure and impure samples, the complex susceptibility
$\chi(\omega)=\chi^{\prime}(\omega)+i\chi^{\prime\prime}(\omega)$
was measured in presence of a moderate static magnetic field
($H_{\rm dc}=2$ kOe) and of a much smaller ac field oscillating at
frequency $\omega$. A two-peaked structure was found in
$\chi^{\prime}(\omega)$ as a function of temperature: the low
temperature peak is frequency-dependent, while the high
temperature one is not. On the basis of transfer matrix
calculations for the static susceptibility of the doped chain, the
low temperature peak was attributed to finite-size
effects.\cite{Bogani_2005} Anyway, the frequency dependence of the
peak for $\chi^{\prime}(\omega)$ and
$\chi^{\prime\prime}(\omega)\ne 0$ remains unexplained. Static and
dynamic susceptibilities with a similar behavior have also been
observed in different cobalt-organic single-chain
magnets,\cite{Li_2006} in presence of a moderate static magnetic
field ($H_{\rm dc}=0.5$ kOe) and of lattice imperfections. The
explanation of such features is thus becoming more pressing, as it
can constitute an important tool for the analysis of the
properties of a whole class of magnetic systems.

In systems with a very large correlation length, like the
one-dimensional Ising model at low temperatures, the introduction
of an external magnetic field can have dramatic consequences. A
static field $H_{\rm dc}$ strongly depresses the correlation
length $\xi$,\cite{Baxter} and this fact, in turn, should also
strongly affect the dynamic susceptibility.
The Glauber dynamics of the infinite Ising chain model in an
external magnetic field was studied some years
ago in order to describe the kinetics of the helix-coil transition
in biopolymers.\cite{Binder} For single-chain magnets, a
theoretical and experimental study was recently performed by
Coulon {\it et al.}\cite{Coulon} focusing on the relaxation time
of the magnetization fluctuations. In addition to a static
magnetic field, their theoretical analysis\cite{Coulon} considered
finite-size effects, relevant in SCMs. A local-equilibrium
approximation was adopted\cite{Coulon} in order to truncate the
infinite hierarchy of kinetic equations for finite open Ising
chains in $H_{\rm dc} \ne 0$. The main advantages of this
approximation, first proposed by Huang\cite{Huang} for infinite
chains, are: (i) it provides the exact steady-state
solution,\cite{Huang} in contrast with the mean field
approximation; (ii) it is valid for any value of the applied field,
in contrast with perturbation methods.\cite{Glauber_1963}

In this work, we develop the theoretical framework necessary to
analyze the ac susceptibility measurements of single-chain magnets
in presence of a static magnetic field $H_{\rm dc}$. We include in
the theory\cite{Coulon} the effect of an oscillating magnetic
field with intensity much smaller than that of the static one,
using a linear response framework. With these theoretical tools,
we can account for the dynamic behavior of CoPhOMe and other
single-chain magnets. We directly compare the calculated behavior
to previous\cite{Bogani_2005} and new experimental measurements of
both the relaxation time and the ac susceptibility in presence of
a static magnetic field $H_{\rm dc}$. In this way, we can
reproduce the temperature and frequency dependence of the ac
susceptibility of SCMs, measured in moderate static fields
($H_{\rm dc} \le 2$ kOe) and in presence of crystal defects and/or
non-magnetic impurities.
For static fields in this range, new data for the relaxation time
of the magnetization of CoPhOMe, $\tau$ versus $H_{\rm dc}$, are
analyzed at low temperature. They are well reproduced by theory,
provided that finite-size effects are included.

The paper is organized as follows. In Section \ref{Model} we
present both the real system and the simplified model that we adopt to
catch the essentials of its stochastic dynamics. In Section
\ref{Statica} we calculate, for pure and doped chain systems, the
temperature dependence of the magnetization and static
susceptibility in presence of a static magnetic field. In
Section \ref{Dinamica}, the theoretical framework for the
calculation of the dynamic susceptibility in presence of a static
magnetic field is first developed for the infinite chain, and then
generalized to an open, finite chain with $N$ spins; next we present
and discuss the approximation of a single dominating contribution, with
characteristic frequency $\Omega_c$, to the dynamic susceptibility.
In Section \ref{Risultati}, the results of our explicit
calculations, performed using parameters suitable for describing
CoPhOMe, are shown and discussed. Finally, the
conclusions are drawn in Section \ref{Conclusioni}, and some
technical details are reported in the Appendices, for the reader's
convenience.

\section{The model}
\label{Model}

The magnetic properties of CoPhOMe are determined by Co(II) ions,
with an Ising character and effective $S=1/2$, and by NITPhOMe
organic radicals, magnetically isotropic and with
$s=1/2$.\cite{Caneschi_2001,Caneschi_2002_C,Caneschi_2002} The
primitive magnetic cell is made up of three cobalt ions and three
radicals. The spins are arranged in a helical structure, with the
helix axis coincident with the crystallographic $c$ axis of the
chain. Although the effective spins of the two types of magnetic
centers have the same value ($1/2$), the gyromagnetic factors are
different. The gyromagnetic factor $g_R$ of the organic radical is
isotropic, while cobalt is strongly anisotropic and $g_{\rm
Co}=g^{\Vert}_{\rm Co} \gg g^{\perp}_{\rm Co}$, with $g_{\rm
Co}\gg g_R$. The Hamiltonian currently used\cite{Caneschi_2002} to
describe a CoPhOMe chain of $N$ spins reads
\begin{eqnarray}
\label{Ising_Co} && \mathcal{H}=-\sum_{l=1}^{N/6}\sum_{m=1}^3 \,
\Bigg\{ J ~\sigma_{l,2m} \Big[\sigma_{l,2m-1} + \sigma_{l,2m+1}
\Big] + \mu_B H  \cr && \Big[ g_{\rm R} \sigma_{l,2m-1}
(\hat{\mathbf{z}}_{2m-1} \cdot \hat{\mathbf{e}}_{\rm H}) + g_{\rm
Co} \sigma_{l,2m} (\hat{\mathbf{z}}_{2m} \cdot
\hat{\mathbf{e}}_{\rm H})\Big] \Bigg\},
\end{eqnarray}
where $\sigma$ is the spin variable, $l$ is the magnetic cell
index and $m$ the site label. The nearest neighbor exchange is
antiferromagnetic and rather strong, $\vert J\vert/k_B \approx 80$
K; $\hat{\mathbf{e}}_{\rm H}$ denotes the direction of the applied
magnetic field $H$; finally, all local axes along which the spins
are aligned, $\hat{\mathbf{z}}_{k}$ ($k=1,\cdots,6$), form the
same angle $\theta \approx 55^{\rm o}$ with the $c$ axis. It
follows that the sublattice magnetizations are not compensated
along $c$, whereas they are compensated within the plane
perpendicular to the chain axis. Thanks to its ferrimagnetic and
quasi-one-dimensional character (the ratio between interchain and
intrachain exchange constants is less than
$10^{-6}$),\cite{Caneschi_2001} CoPhOMe was the first magnetic
molecular compound to display slow relaxation of the magnetization
at low temperatures for $H=0$,\cite{Caneschi_2001} a feature which
was predicted long time ago by Glauber\cite{Glauber_1963} in a
one-dimensional model of Ising spins, coupled by a nearest
neighbor ferromagnetic exchange and interacting with a heat
reservoir, causing them to change their states randomly with time.

In this paper we are primarily concerned with analyzing finite-size
effects on the spin dynamics of CoPhOMe in presence of a
non-negligible static magnetic field. Since model (\ref{Ising_Co})
is too involved, in the following we adopt a simplified model, yet
able to catch the essentials of the dynamic behavior. Namely, we
make the approximation of an open Ising chain with $N$ equal
spins, all with $\sigma=\pm 1$ and the same (isotropic)
gyromagnetic factor $g$, coupled by an effective nearest neighbor
ferromagnetic exchange $J_{\rm I}>0$ and subject to a
time-dependent external magnetic field $H(t)$
\begin{equation}
\label{IsingFM} \mathcal{H}=-\sum_{j=1}^{N} \Bigg[ J_{\rm
I}\sigma_{j}\sigma_{j+1} + g \mu_B H(t)  \sigma_{j} \Bigg].
\end{equation}

\section{Static properties}\label{Statica}

For pure CoPhOMe, static magnetization measurements were
originally performed in single-crystal samples\cite{Caneschi_2001}
with a static magnetic field $H_{\rm dc}$=1 kOe applied parallel
to the chain direction. A strong increase in the quantity $T \cdot
M/H_{\rm dc}$ is found on decreasing $T$ below 100 K, with a
maximum reached around 25 K and a subsequent decrease. Such a
behavior, typical of ferromagnetic and ferrimagnetic systems, was
also observed in the case of poly-crystalline powder samples of
pure\cite{Lascialfari_2003} and Zn(II)-doped
CoPhOMe,\cite{Mariani_2008} at different static magnetic fields
and different concentrations of Zn(II) non-magnetic impurities. It
is also a common feature of all SCMs identified so
far.\cite{Rassegna_SCM}

For an open chain of $N$ spins in an applied field, described by
Eq.~(\ref{IsingFM}), the static properties can be calculated
analytically in terms of the two eigenvalues, $\lambda_0$ and
$\lambda_1$, and of the corresponding eigenvectors of the transfer
matrix.\cite{Baxter,Wortis} Denoting the concentration of quenched
non-magnetic impurities by $c$, the magnetization per spin of the
doped chain is
\begin{equation}
\label{Mstatic} M_{\rm doped}=\sum_{N=1}^{\infty} c^2(1-c)^N~ M_N
\end{equation}
and the static susceptibility per spin is
\begin{equation}
\label{chistatic} \chi_{\rm doped}=\sum_{N=1}^{\infty} c^2(1-c)^N~
\chi_N,
\end{equation}
where the explicit expression for $M_N$ and $\chi_N$ are reported
in Appendix A.

For the infinite chain (i.e. $c=0$ and $N\to \infty$), the static
properties can be calculated analytically\cite{Baxter,Wortis} in
terms only of the larger eigenvalue $\lambda_0$
\begin{equation}
\lambda_0=e^{K}(\cosh h_0 + \sqrt{\sinh^2 h_0 +e^{-4K}}),
\end{equation}
\noindent where $K={{J_I}\over {k_B T}}$ and $h_0={{g \mu_B H_{\rm dc}}\over {k_B T}}$.
The magnetization per spin is
\begin{equation}
\label{Minfinity} M=k_B T~ {1\over
{\lambda_0}}{{\partial\lambda_0}\over {\partial H_{\rm dc}}}=
{{\sinh h_0 }\over {\sqrt{\sinh^2 h_0+e^{-4K}}}}\equiv m_{eq}
\end{equation}
and the susceptibility per spin is
\begin{equation}
\label{chiinfinity} \chi=k_BT~ \Big[
-\dfrac{1}{\lambda_0^2}\Big(\dfrac{\partial \lambda_0}{\partial
H_{\rm dc}}\Big)^2
+\dfrac{1}{\lambda_0}\dfrac{\partial^2\lambda_0}{\partial H_{\rm dc}^2}
\Big].
\end{equation}
\begin{figure}
\includegraphics[width=8cm,bbllx=125pt, bblly=149pt,
bburx=541pt,bbury=544pt,clip=true]{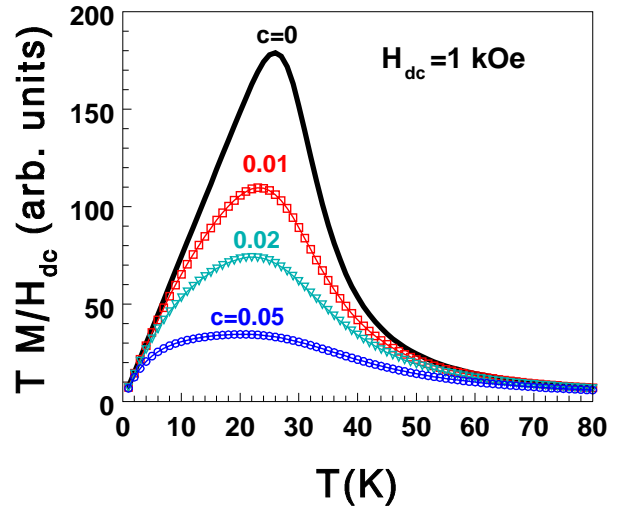} \caption{$T \cdot
M/H_{\bf dc}$ versus $T$ for an Ising chain with $J_I/k_B=80$ K,
in a static magnetic field $H_{\rm dc}=1$ kOe, calculated for
selected values of the concentration $c$ of non-magnetic
impurities (different symbols refer to $c$=0.05, 0.02, 0.01 on
going from bottom to top), and for the pure system ($c=0$, full
line).}\label{Caneschi}
\end{figure}

\begin{figure}
\includegraphics[width=8.5cm,bbllx=107pt, bblly=267pt,
bburx=529pt,bbury=544pt]{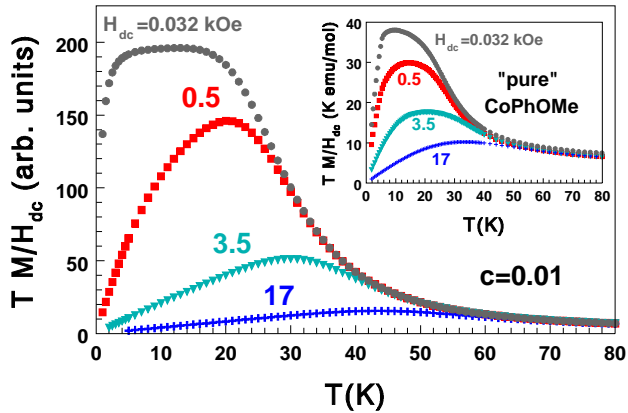}
\caption{(Color online)
$T \cdot M/H_{\rm dc}$ versus $T$ for a doped Ising chain
with $J_I/k_B=80$ K,
calculated for different values of the static applied field, at a
fixed value ($c=0.01$) of the non-magnetic impurities
concentration. Inset: experimental data for a nominally pure
($c=0$) powder sample.}\label{Mstatica}
\end{figure}

\begin{figure*}
\includegraphics[width=16cm,bbllx=107pt, bblly=416pt,
bburx=520pt,bbury=556pt]{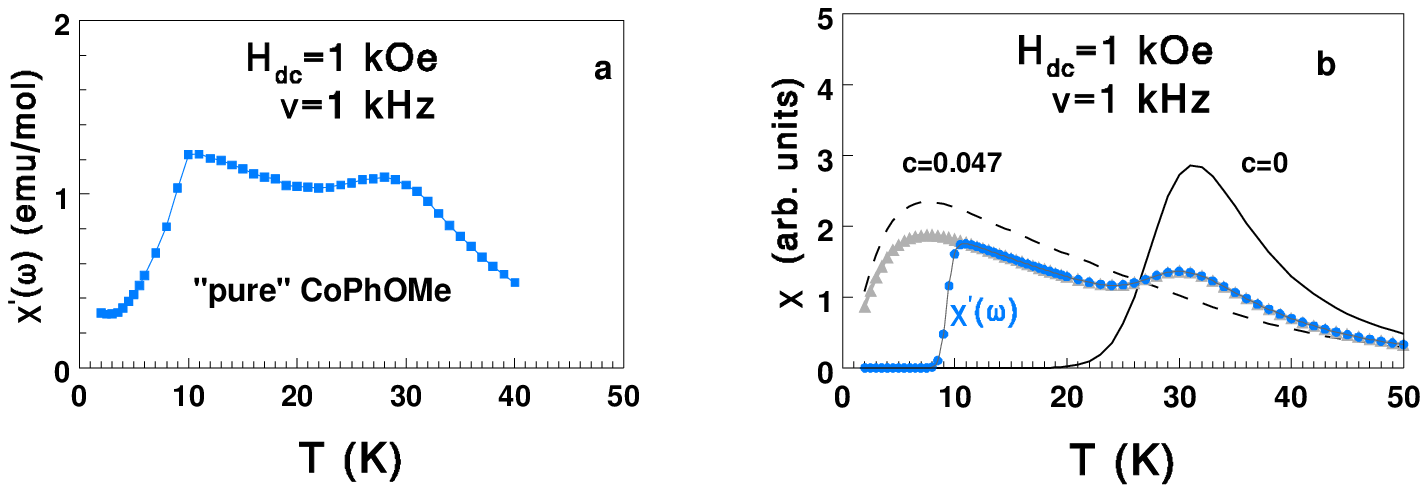}\caption{(Color online) (a)
Real part of the ac susceptibility, $\chi^{\prime}(\omega)$,
(square symbols) measured versus temperature in a nominally pure
CoPhOMe sample for frequency $\nu=1$ KHz and fixed static applied
field $H_{\rm dc}=1$ kOe. (b) Static susceptibility, $\chi$,
calculated with $J_I/k_B=80$ K and $H_{\rm dc}=1$ kOe, for a doped
Ising chain ($c=0.047$, dashed line) and for an infinite chain
($c=0$, full line). The two-peaked structure (light-grey
triangles) is obtained by taking a weighted  average of the two
quantities (80 \% versus 20 \%, respectively). The real part of
the dynamic susceptibility, $\chi^{\prime}(\omega)$, calculated
according to the theory in Section \ref{Dinamica}, is denoted by
blue circles.}\label{Lapostatica}
\end{figure*}

In Fig.~\ref{Caneschi} the calculated quantity $T\cdot M/H_{\rm
dc}$, see Eq.~(\ref{Mstatic}), is reported versus $T$ for
different values of $c$ (including $c=0$) and for a fixed value of
the dc field ($H_{\rm dc}=1$ kOe). In Fig.~\ref{Mstatica}, the
same quantity is reported for different values of the static
applied magnetic field and for a fixed, rather small, value of the
non-magnetic impurities concentration ($c=0.01$). It is worth
observing that, at low fields and low temperatures, the curves
calculated for $c=0.01$ resemble the experimental ones, obtained
by Lascialfari {\it et al.}\cite{Lascialfari_2003} for a powder
sample of nominally pure CoPhOMe (see inset). In contrast,
according to Eq.~(\ref{Minfinity}), the theoretical curves for the
pure system (not shown here) are found to undergo a much stronger
increase at such low fields and temperatures (as it can also be
inferred from the concentration dependence of $T\cdot M/H_{\rm
dc}$ shown in Fig.~\ref{Caneschi}). Previous
suggestions\cite{PRL_2004,Vindigni_2004,Bogani_2005} about the
presence of lattice imperfections, or impurities that limit the
chain size even in pure CoPhOMe, appear thus to be confirmed.

In Fig.~\ref{Lapostatica} we show new experimental data for the
temperature dependence of $\chi^{\prime}(\omega)$, the real part
of the ac susceptibility of a nominally pure CoPhOMe sample,
measured at frequency $\nu=1$ kHz for a fixed value of the static
applied field, $H_{\rm dc}=1$ kOe. For the same field, the
calculated static susceptibility, $\chi$, of the infinite chain
($c=0$, full line) is exponentially vanishing at low temperatures,
see Eq.~(\ref{chiinfinity}), in strong contrast with the diverging
behavior of $\chi$ in zero field. For the doped chain ($c=0.047$,
dashed line), the static susceptibility was calculated using
Eq.~(\ref{chistatic}). It turns out that $\chi_{\rm doped}$,
calculated within the simplified model (\ref{IsingFM}), is in very
good agreement with previous results from an exact transfer matrix
calculation,\cite{Bogani_2005} performed for the more complete
model (\ref{Ising_Co}), as well as with ac susceptibility data in
zinc-doped CoPhOMe.\cite{Bogani_2005} In particular the feature of
a peak gradually shifting to lower temperatures with increasing
$c$, a clear signature of finite-size effects, is recovered (not
shown here) by our simplified model. From Fig.~\ref{Lapostatica}
one sees that the two-peaked feature, experimentally observed at
nonzero field in $\chi^{\prime}(\omega)$ versus $T$, can be well
reproduced by taking a weighted average (light-grey triangles)
between the static susceptibility of the pure chain (black full
line) and that of the doped one (black dashed line). Finally we
mention that also the steep decrease, displayed by the
experimental $\chi^{\prime}(\omega)$ for $T \le 10$ K, can be well
reproduced (blue circles) by taking into account dynamic effects,
at it will be shown in Section \ref{Dinamica}. In summary, the new
data and calculations reported in Fig.~\ref{Lapostatica} not only
are in good agreement with the reported presence of a distribution
of impurities inside the crystal,\cite{Bogani_2005} but they are
also revealing of a non-homogeneous nature of the samples.

A similar two-peaked feature was observed, later, in the real part
of the ac susceptibility of a different cobalt-organic
single-chain magnet, for $H_{\rm dc}=0.5$ kOe.\cite{Li_2006} As in
the case of CoPhOMe, the higher-temperature peak, due to the
infinite chain, did not present any dynamic effect, while the
lower temperature shoulder, related to finite-size effects, was
found to be frequency-dependent. In the following we will show
that the latter feature can be well accounted for by a calculation
of the dynamic susceptibility in the framework of the simplified
model (\ref{IsingFM}).

\section{Dynamic Properties}\label{Dinamica}

In ac magnetic measurements, a small ac drive magnetic field is
superimposed on the dc field, causing a time-dependent moment in
the sample. Therefore we are faced with the theoretical problem of
determining how the kinetic equations of motion for the
time-dependent spin averages of a finite, open chain in zero
magnetic field\cite{DaSilva,Luscombe} are modified by the presence
of magnetic field of the general form
\begin{equation}
\label{HDCAC}
H(t)=H_{\rm dc}+H_1~ e^{-i\omega t}
\end{equation}
i.e. the sum of a static dc field of any intensity, $H_{\rm dc}$, and of
an ac field, oscillating at the angular frequency $\omega$.
In the following we will make use of the reduced fields
$h_0={{g\mu_B H_{\rm dc}}\over {k_B T}}$ and
$h_1={{g\mu_B H_1}\over {k_B T}}$. As the experimental oscillating fields
are usually much smaller than the static field, we will consider the case
$H_1\ll H_{\rm dc}$, which allows us to use the expansion
($h(t)={{g\mu_B H(t)}\over {k_BT}}$)
\begin{equation}\tanh h(t) \approx \tanh h_0 + h_1 ~e^{-i \omega
t}(1-\tanh^2 h_0).\end{equation}
Generally speaking, the susceptibility induced by a field as
in Eq.~(\ref{HDCAC}) will have a real and an imaginary part
\begin{equation}
\chi(\omega)=\chi^{\prime} (\omega) + i \chi^{\prime\prime}(\omega)
\end{equation}
from which the dynamic behavior of molecular materials is usually extracted.
In the following we will thus focus on calculating these experimentally
relevant quantities, which are compared to data in the next section.

Typical values of the frequencies, $\nu={{\omega}\over {2\pi}}$,
used in ac susceptibility
measurements\cite{Caneschi_2001,Bogani_2005} on pure and doped
CoPhOMe, range between 0.1 and 10000 Hz,. In contrast, one has
$\nu \approx$ MHz for proton nuclear magnetic resonance (NMR) and
muon spin relaxation ($\mu$SR)
experiments.\cite{Micotti_2004,Mariani_2007} In the following,
before passing to the comparison with experimental data, we first
examine the properties of an infinite chain (subsection A), then
include finite-size effects (subsection B), and eventually explore
the single-frequency approximation (subsection C).

\subsection{Infinite chain}

Before considering finite-size effects, it is instructive to
calculate first the ac susceptibility of an infinite Ising chain
in presence of a static magnetic field.
The kinetic equation
for the site-independent average $m(t)=\langle \sigma(t) \rangle$
of a spin in the infinite chain is
\begin{equation}
\label{infinito} \tau_0 {{dm(t)}\over {dt}}=-(1-\gamma)m(t)+
\Big[1-\gamma \Gamma_1(t)  \Big]\tanh h(t)
\end{equation}
where $\gamma=\tanh(2K)$ and $\Gamma_1(t)$ is the nearest neighbor
spin-pair time-dependent correlation function. The kinetic
equation for $\Gamma_1(t)$, on its turn, involves higher order
time-dependent correlation functions, so that eventually an
infinite sequence of equations is obtained as a consequence of
$h(t) \ne 0$. In order to truncate this hierarchy, we adopt the
local-equilibrium approximation:\cite{Huang,Coulon} i.e., the
relation holding at equilibrium\cite{Marsh} between the nearest
neighbor correlation function and the magnetization (where
$m_{eq}$ is defined in Eq.~\ref{Minfinity}),
\begin{eqnarray}
\label{localapprox} \Gamma_{1,eq} &=& m_{eq}^2 +\cr &&(1-m_{eq}^2)
{{\cosh h_0-\sqrt{\sinh^2 h_0+e^{-4K}}}\over {\cosh
h_0}+\sqrt{\sinh^2 h_0+e^{-4K}}},
\end{eqnarray}
\noindent  is assumed to hold {\it locally} also at any time $t
\ne 0$. The main advantages of this approximation, first proposed
by Huang\cite{Huang} for an infinite Ising chain model with
Glauber dynamics in $H_{\rm dc}\ne 0$, are: (i) it provides the
exact steady-state solution,\cite{Huang} in contrast with the mean
field approximation, which assumes simply
$\Gamma_{1,eq}=m_{eq}^2$; (ii) it is valid for any value of the
applied field, in contrast with the perturbation method, which
assumes $h_0 \ll 1$.\cite{Glauber_1963}

In this way, a nonlinear equation for $m(t)$ is obtained, where the approximation
of a linear response of the chain applies only to the ac field, $H_1$.
We thus assume $\delta m(t)=m(t)-m_{eq}$,
i.e. small departures of the magnetization from
its equilibrium value $m_{eq}$.
Likewise, we expand $\Gamma_1(t) \approx
\Gamma_1\vert_{eq}+{{d\Gamma_1(t)}\over {dm(t)}}\vert_{eq}~ \delta
m(t)$ and, taking into account that ${{d\Gamma_1(t)}\over {d
m(t)}}\vert_{eq}=2\tanh h_0$ at equilibrium,\cite{Coulon} we
finally obtain a linear non-homogeneous differential equation for
$\delta m(t)$
\begin{eqnarray}
\label{diffinfty} &&\tau_0 {{d~\delta m(t)}\over
{dt}}=-(1-\gamma+2\gamma \tanh^2 h_0)~\delta m(t) \cr &&+h_1~
e^{-i \omega t} (1-\tanh^2 h_0) \Big( 1-\gamma ~\Gamma_{1,eq}
\Big).
\end{eqnarray}

In absence of the ac field ($h_1=0$), one finds an exponential
time decrease for the magnetization fluctuation
\begin{equation}
\delta m(t) = \delta m(t_0)
e^{-\lambda_{\infty}(t_{}-t_{0})/\tau_0}.
\end{equation}
Thus, for the infinite Ising chain, there is a single relaxation
time, $\tau_{\infty}$, related to the a-dimensional parameter
$\lambda_{\infty}$ by
\begin{equation}
\label{lambdainfty} \lambda_{\infty}={{\tau_0}\over
{\tau_{\infty}}}=1-\gamma+2\gamma\tanh^2 h_0.
\end{equation}
Notice that, for $h_0 \to 0$, Glauber's result\cite{Glauber_1963}
of an exponentially diverging relaxation time at low temperatures
($\tau_{\infty}\approx {1\over 2} \tau_0 e^{4K}$ for $k_B T \ll
J_I$) is correctly recovered, while for $H_{\rm dc} \ne 0$ the
relaxation time of the infinite chain does not diverge any more.

In presence of the ac field ($h_1 \ne 0$), the general
solution of Eq.~(\ref{diffinfty}) is
\begin{eqnarray}
&&\delta m(t) =\delta m(t_0)
e^{-\lambda_{\infty}(t_{}-t_{0})/\tau_0}+h_1~(1-\tanh^2 h_0)\cr &&
\times (1-\gamma ~\Gamma_{1,eq}) \int_{t_0}^t~e^{-i \omega
t^{\prime}}
e^{-\lambda_{\infty}(t^{}-t^{\prime})/\tau_0}~dt^{\prime}.
\end{eqnarray}
Taking into account that $\lambda_{\infty}\ne 0$, one can safely
let $t_{0} \to -\infty$ in order to find a solution that does not
depend on the initial conditions\cite{Glauber_1963,BreyPrados}
\begin{equation}
\delta m(t) =(1-\tanh^2 h_0)( 1-\gamma
\Gamma_{1,eq}){{h_1 e^{-i \omega t}}\over
{\lambda_{\infty}-i\omega \tau_0 }} .
\end{equation}
The fluctuation of the total magnetization of the infinite Ising
chain is obtained summing over all the $N$ spins and letting
$N \to \infty$
\begin{equation}
\delta \langle M(t) \rangle =N g \mu_B \delta \langle \sigma (t)
\rangle=\chi(\omega) H_1 e^{-i \omega t}, \end{equation} so that
the dynamic susceptibility, $\chi(\omega)$, is
\begin{eqnarray}
\label{chi_omega_Glauber}
&&\chi(\omega)={{g^2 \mu_B^2 N}\over {k_B T}}~  (1-\tanh^2 h_0)(
1-\gamma \Gamma_{1,eq})\cr&& \times {1\over
{\lambda_{\infty}-i\omega \tau_0 }}={{\chi_{\infty}}\over
{1-i\omega {{\tau_0}\over {\lambda_{\infty}}} }}
\end{eqnarray}
\noindent where $\chi_{\infty}$ denotes the static susceptibility
of the infinite Ising chain for $H_{\rm dc} \ne 0$
\begin{equation}
\label{susc_DC_infty}
\chi_{\infty}={{g^2 \mu_B^2 N}\over {k_B T}}~{{(1-\tanh^2 h_0)(
1-\gamma \Gamma_{1,eq})}\over {1-\gamma+2\gamma\tanh^2 h_0}}.
\end{equation}

For $h_0 \to 0$, Glauber's result\cite{Glauber_1963}
\begin{equation}
\label{chiGlauber} \chi(\omega)=
{{g^2 \mu_B^2 N}\over {k_B T}}~ {{ 1+\eta}\over {1-\eta}}~
{1\over {1-i \omega~{{\tau_0}\over {1-\gamma}} }}
\end{equation}
is correctly recovered, by taking into account that
$\lambda_{\infty} \to 1 -\gamma$, $\Gamma_{1,eq} \to \eta=\tanh K$,
and $\gamma={{2\eta}\over {1+\eta^2}}$.

\subsection{Finite chain}
In the case of an open Ising chain with a finite number $N$ of
spins, the lack of translational invariance leads to $N$
kinetic equations for the $N$ site-dependent spin averages $\delta
\langle \sigma_p(t) \rangle$
($p=1,\cdots,N$).\cite{Luscombe,DaSilva,Coulon}
As in the case of the infinite chain, we can introduce
the local-equilibrium approximation\cite{Huang,Coulon} in order
to truncate the infinite sequence of equations for the
higher-order time-dependent spin correlation function. Next we
perform the linearization of the kinetic equations, in the hypothesis of a
linear response to the ac magnetic field. We then obtain
a set of $N$ linear differential equations,
which can be written in matrix form
\begin{equation}
\label{eqmatrixform} \tau_0 {{d{\bf \Sigma}}\over {dt}}
=-{\bf Y} \cdot {\bf \Sigma}+h_1
e^{-i\omega t}(1-\tanh^2 h_0) {\bf \Psi}.
\end{equation}
where ${\bf \Sigma}$ and ${\bf \Psi}$ are $N\times 1$ vectors
containing the spin fluctuations and the non-homogeneous terms,
respectively (see Appendix B for details). ${\bf Y}$ is a real,
symmetric, tridiagonal, $N\times N$ matrix, with nonzero
a-dimensional eigenvalues $\lambda_j$ ($j=1,...,N$), while ${\bf
\Phi}^{(\lambda_j)}$ are the corresponding $N\times 1$
eigenvectors. In the limiting case $h_{0} \to 0$, the numerical
solutions for $\lambda_j$ coincide with the ones
obtained\cite{Luscombe,DaSilva} in the framework of a finite-size
scaling calculation of the Glauber dynamics in a zero static
field. In particular, the low-temperature expansion ($k_B T \ll
J_I$) for the eigenvalue of a finite open chain with $N$ spins is
$\lambda_1(H_{\rm dc}=0)\approx {2\over {N-1}} e^{{{-2J_I}\over
{k_BT}}}$,\cite{DaSilva} to be compared with
$\lambda_{\infty}(H_{\rm dc}=0)\approx 2 e^{{{-4J_I}\over
{k_BT}}}$ for the eigenvalue of the infinite
chain.\cite{Glauber_1963}

We are interested in the long time behavior of the system,
characterized for being independent on the initial condition.
Thus, solving Eq.~(\ref{eqmatrixform}) by the method
of eigenfunctions\cite{vanKampen} and letting $t_0 \to
-\infty$,\cite{Glauber_1963,BreyPrados} we obtain for the
fluctuation of a single-spin average ($p=1,\cdots,N$)
\begin{eqnarray}
\delta \langle \sigma_p (t)\rangle&=& h_1 e^{-i \omega t}
(1-\tanh^2 h_0)\cr &\times& \sum_{j=1}^N {1\over
{\lambda_j-i\omega\tau_0}} \Phi_p^{(\lambda_j)} \Big(\sum_{m=1}^N
\Phi_m^{(\lambda_j)}\Psi_m^{}\Big)
\end{eqnarray}
The fluctuation of the magnetization of the finite open Ising
chain is obtained summing over the $N$ spins
\begin{equation}
\delta\langle M_N (t) \rangle=g\mu_B \sum_{p=1}^N \delta\langle
\sigma_p (t) \rangle = \chi_N(\omega) H_1 e^{-i \omega t}
\end{equation}
The dynamic susceptibility $\chi_N(\omega)$ takes the form
\begin{equation}
\label{chi_omega}\chi_N(\omega)={{g^2 \mu_B^2}\over {k_B T}} \sum_{j=1}^N
{{\Omega_j^2+i\omega\Omega_j}\over {\Omega_j^2+\omega^2}}
A_j(\lambda_j,T,H_{\rm dc})
\end{equation}
\noindent where the angular frequencies are
\begin{equation}
\label{dimensfreq} \Omega_j={{\lambda_j}\over {\tau_0}}
\end{equation}
\noindent and the corresponding frequency weights are
\begin{eqnarray}\label{weights}
A_j(\lambda_j,T,H_{\rm dc})&=& {{1-\tanh^2 h_0}\over {\lambda_j}}\sum_{p=1}^N
\Phi_p^{(\lambda_j)} \cr &\times& \Big(\sum_{m=1}^N
\Phi_m^{(\lambda_j)}\Psi_m^{}\Big).
\end{eqnarray}
The angular frequencies (\ref{dimensfreq}) are expressed in terms of the a-dimensional
eigenvalues, $\lambda_j$, of the matrix ${\bf Y}$ and of the
characteristic time, $\tau_0$, for the spin flip of an isolated
spin. The latter is a free parameter which is expected to depend, in general,
on the intensity of the applied magnetic field.\cite{Coulon}  The general expression
for the dynamic susceptibility per spin of a doped chain is thus
\begin{equation}
\label{chiAC} \chi(\omega)=\sum_{N=1}^{\infty} c^2(1-c)^N~
\chi_N(\omega).
\end{equation}

\begin{figure*}
\includegraphics[width=16cm,bbllx=19pt, bblly=116pt,
bburx=580pt,bbury=640pt]{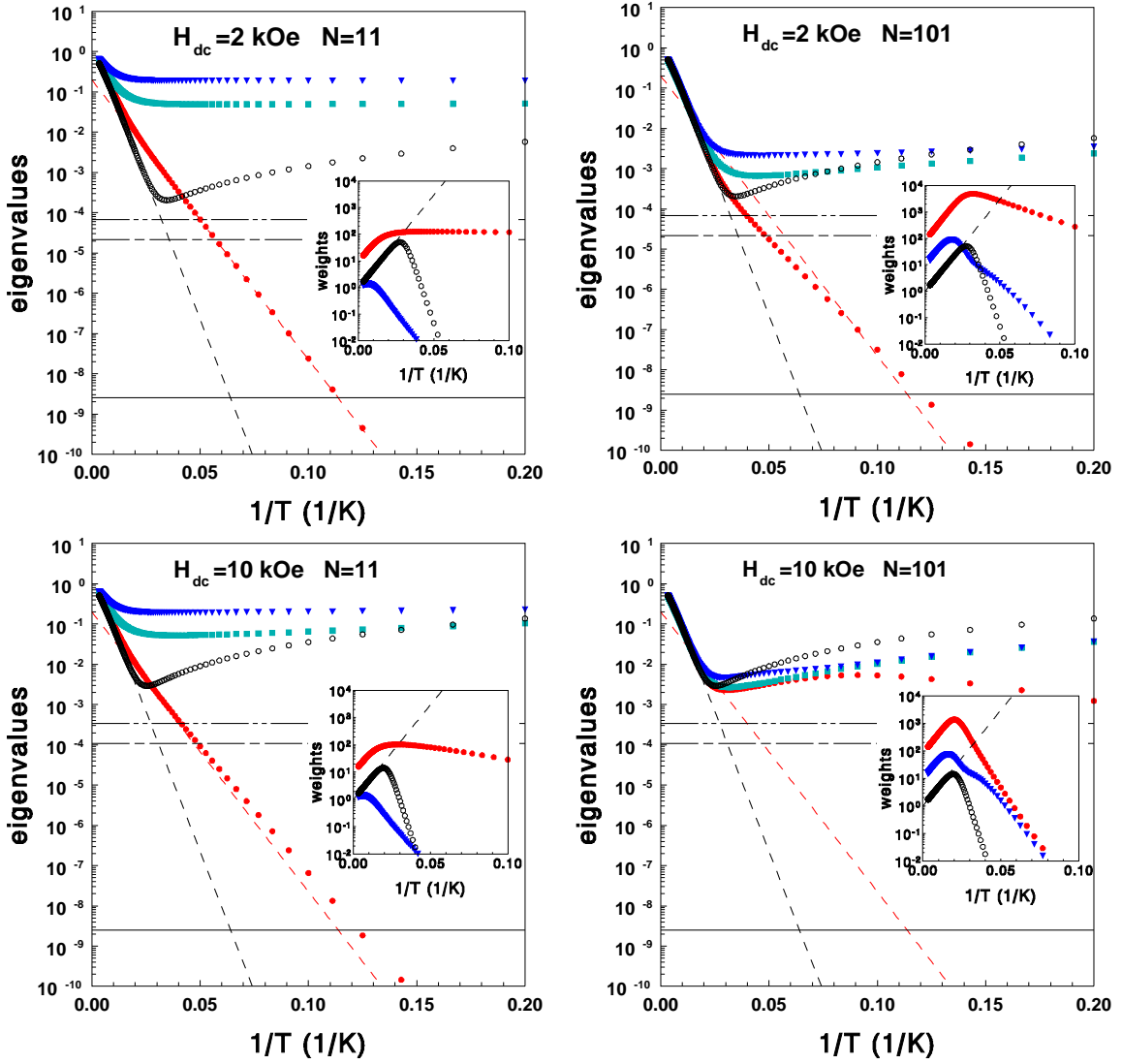} \caption{(Color online)
Temperature dependence of the first three eigenvalues
$\lambda_j=\Omega_j \tau_0$ ($j=1$, red full circles; $j=2$, green
full squares; $j=3$, blue full triangles) of an open Ising chain,
calculated for a finite number of spins ($N$=11 and $N=101$) in
presence of a nonzero static field ($H_{\rm dc}=2$ kOe and $H_{\rm
dc}=10$ kOe). For comparison, the eigenvalue $\lambda_{\infty}$
(black open circles) of an infinite chain in nonzero field is also
reported. Dashed lines denote the zero-field, low-temperature
expansions for the smallest eigenvalue of a finite chain,
$\lambda_1(H_{\rm dc}=0)\approx {2\over {N-1}} e^{-{{2J_I}\over
{k_BT}}}$,\cite{DaSilva} and for the eigenvalue of an infinite
chain, $\lambda_{\infty}(H_{\rm dc}=0)\approx 2 e^{-{{4J_I}\over
{k_BT}}}$.\cite{Glauber_1963} The horizontal lines, from bottom to
top, denote the quantity $2\pi\nu\tau_0$, calculated for three
different frequencies: $\nu=1$ kHz, $\nu$(MHz)$=4.26 \cdot H_{\rm
dc}$(kOe), and $\nu$(MHz)$=13.55\cdot H_{\rm dc}$(kOe), typically
used in ac susceptibility, proton NMR, and $\mu$SR measurements,
respectively. Insets: calculated temperature dependence of the
frequency weights, $A_j(\lambda_j,T,H_{\rm dc})$ with $j=1$ and
$j=3$ (red full circles and blue full triangles, respectively);
for odd $N$, the weights of the even modes are zero\cite{Coulon}.
The curves denoted by black open circles and black dashed lines
represent the frequency weights of an infinite chain in $H_{\rm
dc} \ne 0$ and in $H_{\rm dc}=0$, respectively. The parameters
used for the calculations were $J_I/k_B=80$ K, $g=2$, and
$\tau_0=4\cdot 10^{-13}$ s.}\label{quadrupla}
\end{figure*}

\begin{figure*}
\includegraphics[width=16cm,bbllx=56pt, bblly=269pt,
bburx=527pt,bbury=568pt]{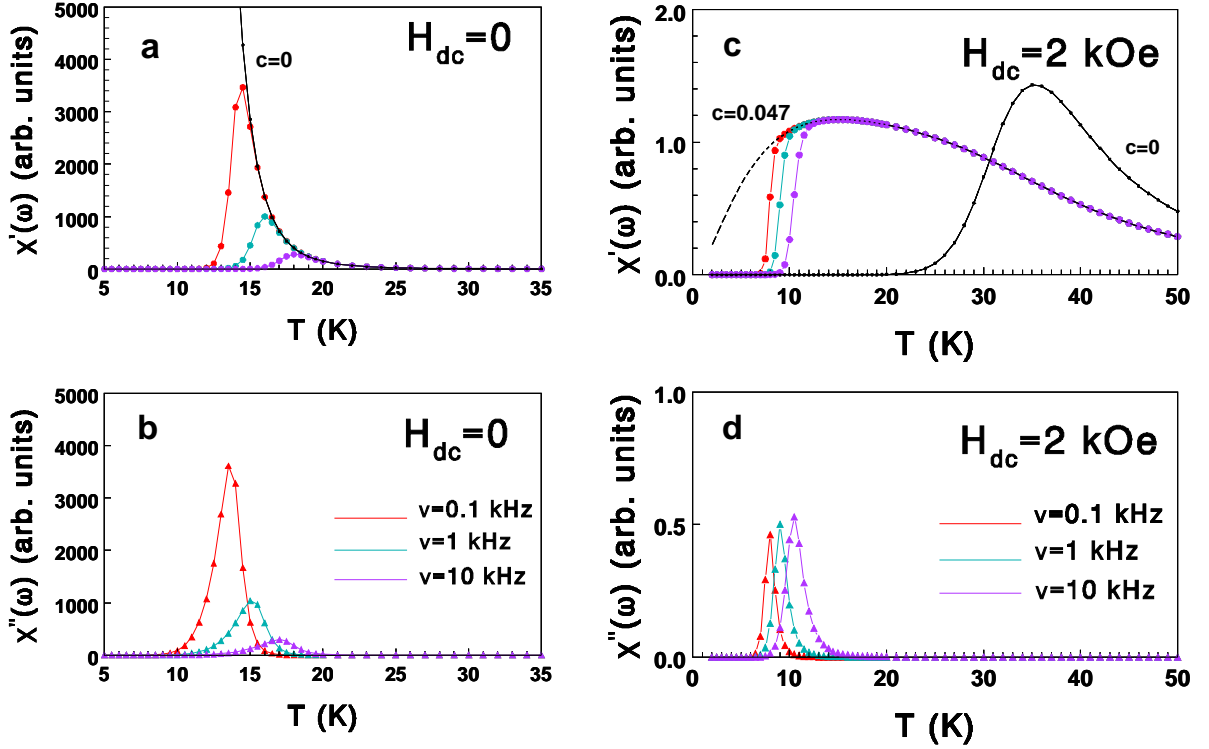} \caption{(Color online)
Temperature dependence of the calculated dynamic susceptibility
for an infinite ($c=0$) ferromagnetic Ising chain in zero static
field $H_{\rm dc}=0$ (figures a and b), and of a doped ($c=0.047$)
Ising chain in nonzero static field $H_{\rm dc}=2$ kOe (figures c
and d). The different curves refer to different frequencies
($\nu=0.1$, 1, and 10 kHz on going from left to right) of the ac
magnetic field. The calculated static susceptibility of the
infinite chain in $H_{\rm dc}=0$, is denoted by the full black
line: it diverges exponentially as $T$ is decreased to 0, see (a),
while in $H_{\rm dc}\ne 0$ it goes through a peak, and then
vanishes exponentially, see (c). The calculated static
susceptibility of the doped ($c=0.047$) chain in nonzero ($H_{\rm
dc}=2$ kOe) field is denoted  by the black dashed line, see
(c).}\label{CaneLapoAC}
\end{figure*}

\begin{figure*}
\includegraphics[width=16cm,bbllx=60pt, bblly=290pt,
bburx=520pt,bbury=580pt]{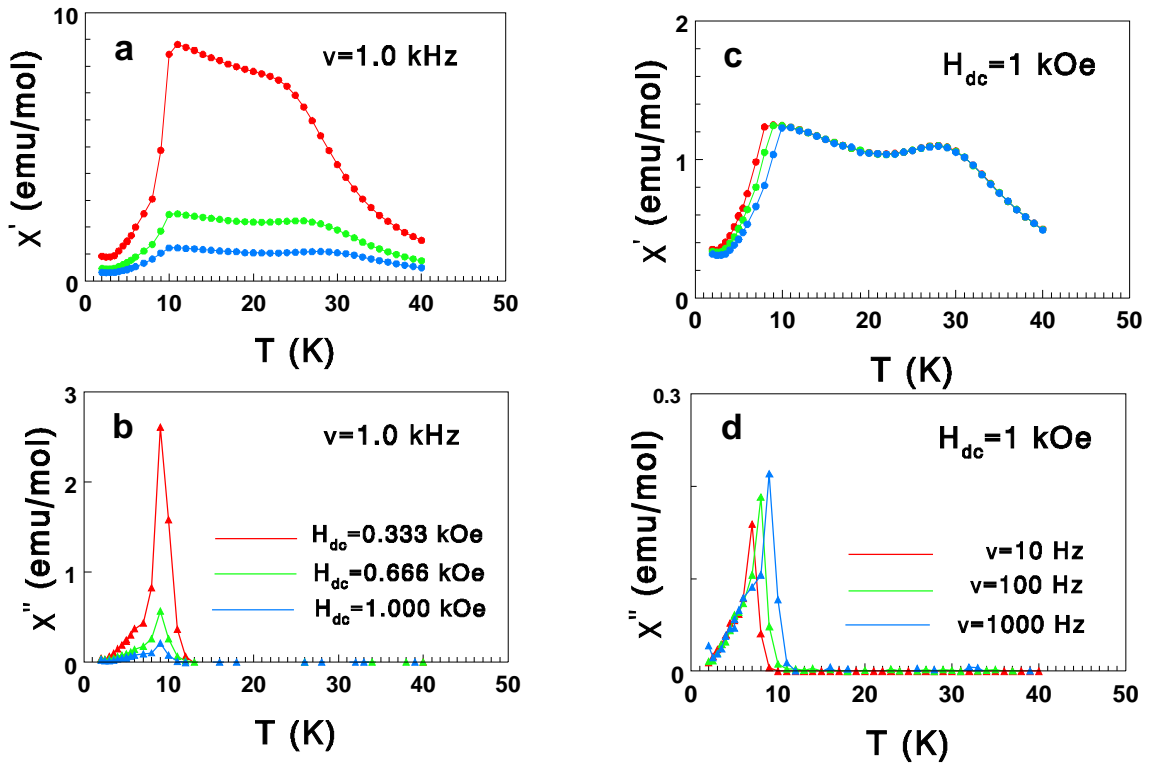} \caption{(Color online)
Temperature dependence of the measured ac susceptibility of
CoPhOMe. The data refer to the same sample as in
Fig.~\ref{Lapostatica}(a). In (a) and (b), the frequency was fixed
($\nu=1$ kHz) and the dc field was varied ($H_{\rm dc}=0.333$,
0.666, and 1 kOe on going from top to bottom). In (c) and (d), the
dc field was fixed ($H_{\rm dc}=1$ kOe) and the frequency was
varied ($\nu=10$, 100, and 1000 Hz on going from left to
right).}\label{figure6_expt}
\end{figure*}

\begin{figure*}
\includegraphics[width=16cm,bbllx=60pt, bblly=290pt,
bburx=520pt,bbury=580pt]{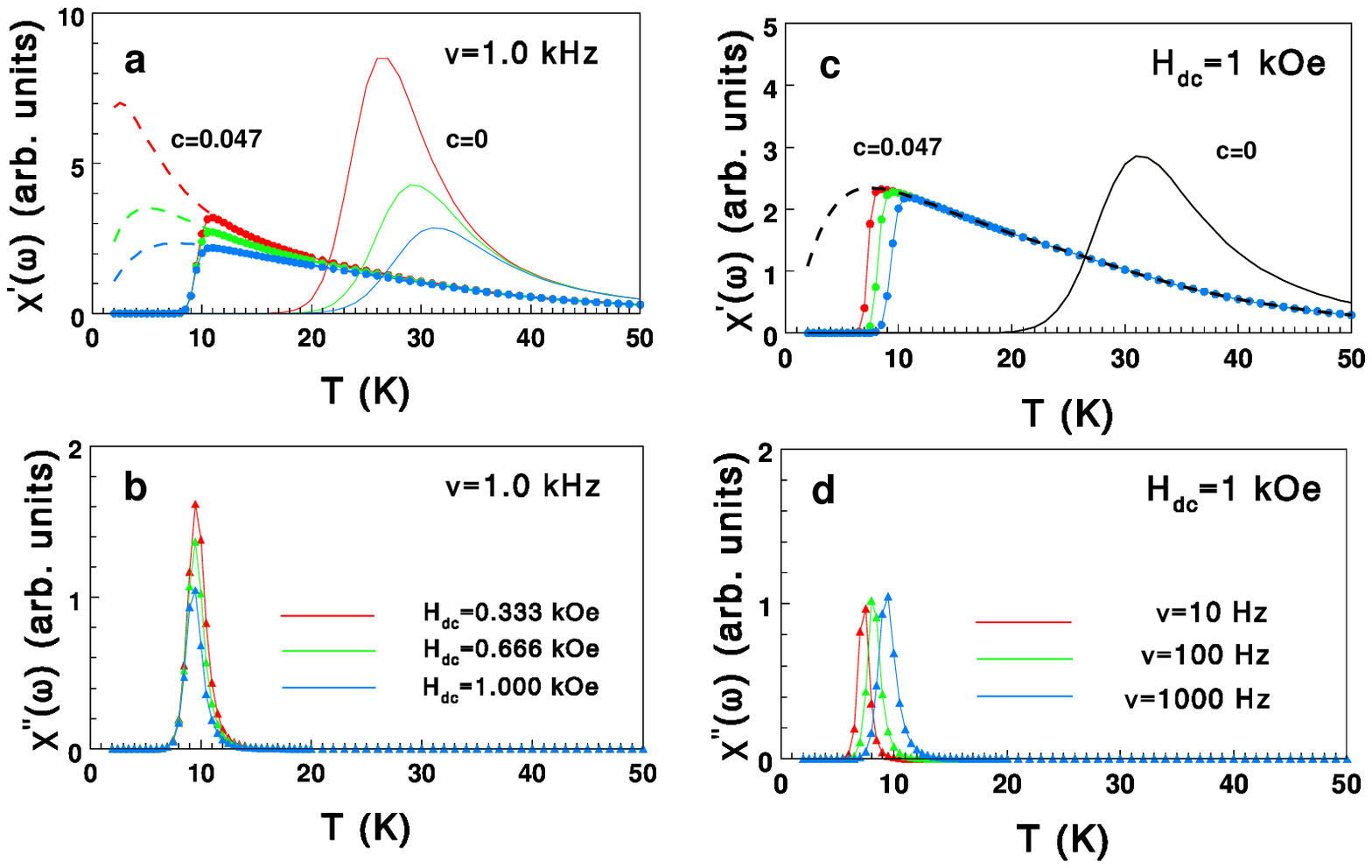} \caption{(Color
online) Temperature dependence of the dynamic susceptibility
of CoPhOMe, calculated for a doped chain ($c=0.047$) in
the single-frequency approximation, see Eq.~(\ref{chiACsingle}) and
Eq.~(\ref{chistatic}), for the same values of frequency and dc field
as in Fig.~\ref{figure6_expt}. Thin dashed lines denote the static
susceptibility of the doped system ($c=0.047$) versus $T$.
Thin solid lines denote the dynamic susceptibility of the
infinite chain ($c=0$): notice the absence of any frequency dependence
for fixed $H_{\rm dc}=1$ kOe, see (c), at the chosen values of $\nu$.
}\label{figure6_theo}
\end{figure*}

\subsection{Single-frequency approximation}

Let us start deriving a quite general sum rule for the frequency
weights, which is readily obtained letting zero frequency
($\omega=0$) in Eq.~(\ref{chi_omega})
\begin{equation}
\sum_{j=1}^N A_j(\lambda_j,T,H_{\rm dc})= {{k_B T}\over {g^2
\mu_B^2}}~\chi_N \end{equation} where $\chi_N$, the static
susceptibility of a finite open Ising chain with $N$ spins,
subject to the dc field $H_{\rm dc}$, is given by
Eq.~(\ref{chiN}). Next we observe that, in the approximation of a
single characteristic frequency, $\Omega_c$, dominating the
relaxation of the magnetization fluctuations, the equation
(\ref{chi_omega}) for the dynamic susceptibility of a finite,
open, Ising chain with $N$ spins assumes the simple form
\begin{equation}
\label{chi_omega_single} \chi_N(\omega) \approx \chi_N~
{{\Omega_c^2+i\omega\Omega_c}\over {\Omega_c^2+\omega^2}}.
\end{equation}
In principle, the characteristic frequency is not necessarily
$\Omega_c=\lambda_1/\tau_0$, i.e. related to the smallest
eigenvalue of ${\bf Y}$ . Rather, in order to determine the
dominating frequency, the temperature dependence of the frequency
weights must be taken into account.

In the single-frequency approximation, the dynamic susceptibility per
spin of a doped chain also simplifies greatly
\begin{equation}
\label{chiACsingle} \chi(\omega) \approx \chi_{\rm doped}~
{{\Omega_c^2+i\omega\Omega_c}\over {\Omega_c^2+\omega^2}},
\end{equation}
\noindent where $\chi_{\rm doped}$ is given by
Eq.~(\ref{chistatic}).

Finally we observe that, using the fluctuation-dissipation
theorem, the linear response ${\cal S}_N(\omega)$ of a finite open
Ising chain with $N$ spins can be expressed as a weighted sum of
$N$ Lorentzian functions centered at zero frequency with widths
$\Omega_j$
\begin{eqnarray}
\label{response} {\cal S}_N(\omega)&=&{{2k_B T}\over {\omega}}
\chi^{\prime\prime}_N(\omega)\cr &=& 2~g^2 \mu_B^2
 \sum_{j=1}^N {{ \Omega_j}\over {\Omega_j^2+\omega^2}}~
A_j(\lambda_j,T,H_{\rm dc}) .
\end{eqnarray}
\noindent When the approximation of a single dominating frequency
(\ref{chi_omega_single}) holds, the linear response is simply the
product of $T\chi_N$ and of a single Lorentzian function, centered
at zero frequency, with width equal to the characteristic
frequency $\Omega_c$
\begin{equation}
\label{response_single} {\cal S}_N(\omega)={{2k_B T}\over
{\omega}} \chi^{\prime\prime}_N(\omega) \approx 2k_B T\chi_N
~{{\Omega_c}\over {\Omega_c^2+\omega^2}} .
\end{equation}
\noindent It is worth noticing that expressions quite similar to
Eqs.~(\ref{response}) and (\ref{response_single}) were obtained
for the spectrum of fluctuations of a cluster magnetization by
Santini {\it et al.},\cite{Santini} in the framework of an exact
calculation of the energy levels of three important classes of
magnetic molecules in contact with a phonon heat bath: namely,
antiferromagnetic rings, grids, and nanomagnets. Moreover, Bianchi
{\it et al.}\cite{Bianchi} recently showed that, while for
antiferromagnetic homometallic rings the approximation of a single
dominating frequency is valid, it does not hold for heterometallic
rings,\cite{Santini} due to the presence of inequivalent ions
which prevent mapping local-spin correlations with the
corresponding total-spin ones.

\section{Results}\label{Risultati}

In this section we compare the theoretical results derived in \ref{Dinamica}
with experimental ac susceptibility data,
obtained in nominally pure and zinc-doped CoPhOMe
for different values of $H_{\rm dc}$ and of the frequency of the oscillating field.
Also, new experimental data for the relaxation time of the magnetization,
measured as a function of the static magnetic field at fixed temperature,
will be discussed and compared with theoretical calculations.

Let us first provide evidence for the correctness of the single
frequency approximation, Eq.~(\ref{chiACsingle}),
by showing the temperature dependence of
the eigenvalues and weights of a finite, open, Ising chain in an
applied dc field. The value of the exchange constant we assumed for the
calculations, $J_I/k_B=80$ K, is known\cite{Vindigni_2004} to provide
the correct temperature dependence for the static and dynamic properties
of CoPhOMe. The characteristic time for the spin flip of an isolated spin,
$\tau_0$, was left as a free parameter (see later on). Moreover
the field dependence of $\tau_0$, though expected in principle,\cite{Coulon}
was neglected for the sake of simplicity.

In Fig.~\ref{quadrupla} some of the $N$ calculated a-dimensional
eigenvalues, $\lambda_j$, of the real tridiagonal
matrix ${\bf Y}$, defined in Eq.~(\ref{eqmatrixform})
and Eq.~(\ref{matrix_elements_Y}), are reported as a function
of inverse temperature for different values of the number of
spins, $N$, and of the applied static field, $H_{\rm dc}$. For the
sake of comparison, also the temperature dependence of the
eigenvalue $\lambda_{\infty}$, see Eq.~(\ref{lambdainfty}), of an
infinite chain in the same field is shown.
Except for the case of long chains and strong fields, at
sufficiently low temperatures a single mode dominates
the low-frequency dynamics of a finite, open, Ising chain with $N$
spins: namely, the mode with characteristic frequency
$\Omega_c=\lambda_1/\tau_0$, where $\lambda_1$ is the smallest
eigenvalue of ${\bf Y}$. The temperature dependence
of the frequency weights corresponding to the various modes is
displayed in the insets of Fig.~\ref{quadrupla}: for not too long
chains and not too strong fields, one can see that the frequency
related to the smallest eigenvalue $\lambda_1$ has
the strongest weight.

\subsection{Ac susceptibility}

Some years ago, the ac susceptibility of CoPhOMe was measured
versus $T$ in single crystals, both for $H_{\rm dc}=0$ Oe in a nominally pure
sample\cite{Caneschi_2001} and for $H_{\rm dc}=2$ kOe in a doped one.\cite{Bogani_2005}
In this work, we present some new data for $\chi(\omega)$ versus $T$. Data were
obtained, using an homemade ac probe and a Cryogenics magnetometer,
on a collection of nominally pure single crystals of CoPhOMe. The
crystals were all aligned with the chain axis along the magnetic field direction.
The frequencies in an ac susceptibility experiment typically
range between 0.1 and 10 kHz, and we can safely adopt the single-frequency
approximation in order to account for the temperature
dependence of $\chi(\omega)$ in moderate fields ($H_{\rm dc}\le 2$ kOe).
Additionally the oscillating fields used were always below 8 Oe, and the condition
$H_{\rm dc} \gg H_1$ is also fulfilled.

\begin{figure*}
\includegraphics[width=16cm,bbllx=80pt, bblly=418pt,
bburx=522pt,bbury=594pt]{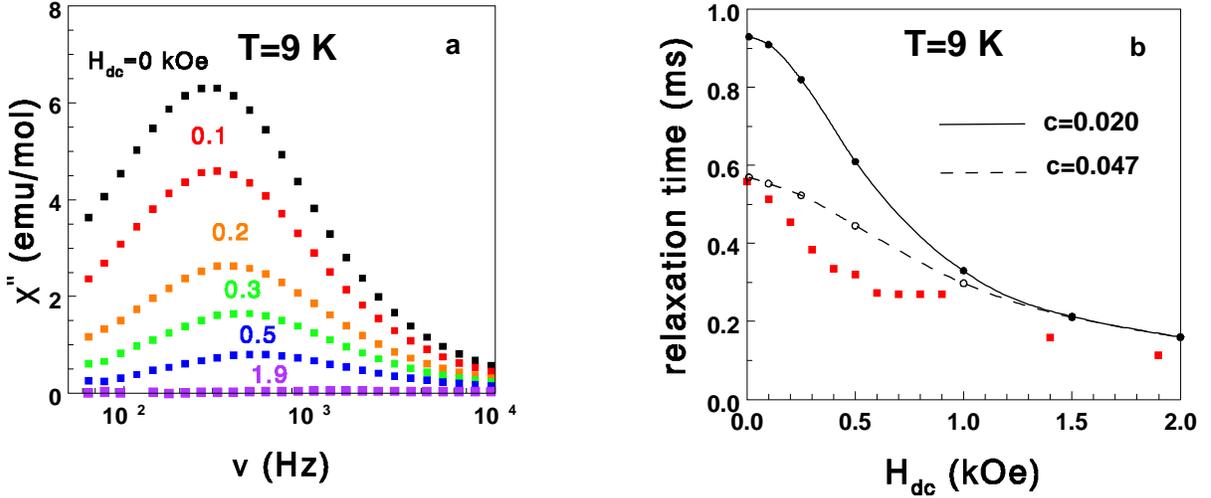} \caption{(Color online) (a)
Frequency dependence of the imaginary part of the ac
susceptibility, $\chi^{\prime\prime}$, measured in CoPhOMe at
fixed temperature $T=9$ K for selected values of the static
magnetic field. (b) Field dependence of the relaxation time $\tau$
of the magnetization of CoPhOMe as deduced from the peak position
of $\chi^{\prime\prime}$ versus frequency at $T=9$ K. Squares
denote experimental results; the lines connect theoretical values
calculated for a doped chain in the approximation of a single
dominating frequency (see Eq.~\ref{chiACsingle} and
Eq.~\ref{chistatic}), using $J_I/k_B=80$ K, $g=2$, and $\tau_0=4
\cdot 10^{-13}$ s.} \label{9K}
\end{figure*}
\begin{figure*}
\includegraphics[width=16cm,bbllx=80pt, bblly=418pt,
bburx=522pt,bbury=594pt]{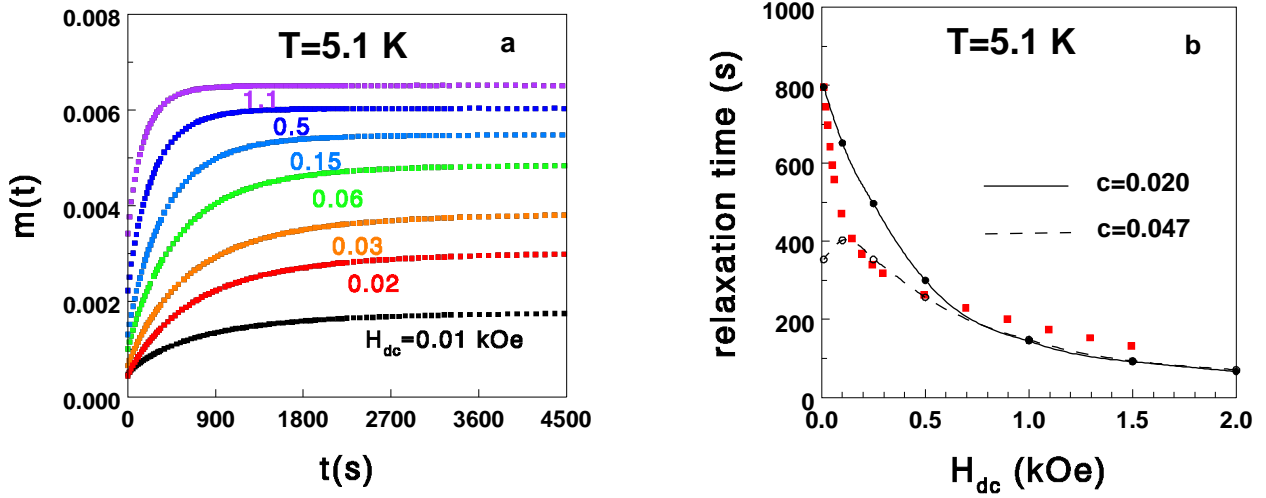} \caption{(Color online) (a)
Time dependence of the magnetization $m(t)$, measured in CoPhOMe
at fixed temperature $T=5.1$ K  for selected values of the static
magnetic field. (b) Field dependence of the relaxation time $\tau$
of the magnetization of CoPhOMe. Squares denote experimental
results obtained by an exponential fit of $m(t)$; the line
connects theoretical values determined as in Fig. 7b, but for
$T=5.1$ K. }\label{5K}
\end{figure*}

In Fig.~\ref{CaneLapoAC}(a)(b) we show the temperature dependence
of the real and imaginary part of the dynamic susceptibility
calculated for an infinite chain with $H_{\rm dc}=0$, while
analogous quantities calculated for a zinc-doped chain ($c=0.047$)
in $H_{\rm dc}=2$ kOe are reported in  Fig.~\ref{CaneLapoAC}(c)(d).

In the pure system ($c=0$) with $H_{\rm dc}=0$, see
Eq.~(\ref{chiGlauber}), a single, resonating peak is found both
for $\chi^{\prime}(\omega)$ and $\chi^{\prime\prime}(\omega)$
versus $T$, whose position gradually shifts to higher temperature
with increasing the frequency of the tiny oscillating field, see
Fig.~\ref{CaneLapoAC}(a)(b). The phenomenon can be
interpreted\cite{BreyPrados,PiniRettori,Vindigni_2009} as a
manifestation of stochastic resonance in a set of coupled bistable
systems: i.e. there is an optimal value of noise, for which the
response of the dynamic system to the driving field is maximum. In
a ferromagnetic or ferrimagnetic chain, the role of stochastic
noise is played by thermal fluctuations, and a resonance peak
occurs when the deterministic time scale of the oscillating
magnetic field matches the statistical time scale associated with
the spontaneous decay of the net magnetization.

For the doped system in $H_{\rm dc}=2$ kOe, we find that a
frequency-dependent peak in the calculated $\chi^{\prime}(\omega)$ and
$\chi^{\prime\prime}(\omega)$ versus $T$ (the
colored curves in Fig.~\ref{CaneLapoAC}(c)(d)) develops at substantially
lower temperatures with respect to the peak in the static
susceptibility of an infinite chain (the full black curve in
Fig.~\ref{CaneLapoAC}(c), see Eq.~\ref{susc_DC_infty}). This can be
easily understood by looking at Fig.~\ref{quadrupla}: for a
finite, open Ising chain in a moderate field, the fulfillment of
the resonance condition ($\omega=\Omega_c$ in
Eq.~\ref{chiACsingle}), occurs at low temperatures, as signaled by
the crossing between the full horizontal line (which represents a
typical value, 1 kHz, of the frequency $\nu$ in an ac
susceptibility measurement) and the curve (red full circles)
representing the $T$ dependence of the smallest frequency mode, $\lambda_1$.
In contrast, the crossing does not occur in the case of an infinite
chain (black open circles) subject to the same dc field: i.e. the relaxation rate of
the infinite chain does not fulfill the resonance condition
($\omega \tau_0=\lambda_{\infty}$ in Eq.~\ref{chi_omega_Glauber})
at the low frequencies involved in the ac susceptibility
experiment: thus, the infinite chain does not present any dynamic
response. As already observed in Sect. III, nominally pure samples
are non-homogeneous and consequently, at low temperatures, only
the regions with dilute chains contribute to the dynamic
properties in a significant way. This explains the experimental
results\cite{Caneschi_2001,Vindigni_2005,Caneschi_2002,PRL_2004,Vindigni_2004}
in which the measured relaxation rate, for $H$=0, was always
found to follow the modified Arrhenius law with a halved energy barrier,
$\tau(T)=\tau_0(\bar{L})~e^{{{2J_I}\over
{k_BT}}}$, where $\bar{L}$ is the mean length of a chain.

In Fig.~\ref{figure6_expt} we present new experimental data for
the temperature dependence of the ac susceptibility of CoPhOMe,
measured for a fixed value of the frequency ($\nu=1$ kHz) at
different values of the static magnetic field ($H_{\rm dc} \le 1$
kOe), see (a)(b), and for a fixed value of the dc field ($H_{\rm
dc}=1$ kOe) at different values of the frequency ($\nu \le 1$
kHz). In Fig.~\ref{figure6_theo} we plot the real and imaginary
part of the dynamic susceptibility, calculated using
Eqs.~(\ref{chiACsingle}) and  (\ref{chistatic}) for a doped chain
($c=0.047$) at just the same values of frequency and dc field as
in Fig.~\ref{figure6_expt}. For the sake of comparison, also the
dynamic susceptibility of the infinite chain ($c=0$) was
calculated: for fixed $H_{\rm dc}=1$ kOe, at the chosen values of
$\nu$, this quantity was not found to display any frequency
dependence (see the black line in Fig.~\ref{figure6_theo}(c)), in
fine agreement with experimental observations (see the higher
temperature peak in Fig.~\ref{figure6_expt}(c)). Thus we can
conclude that, for the moderate values of $\nu$ and $H_{\rm dc}$
exploited in the present ac susceptibility experiments: (i) the
frequency dependence of $\chi(\omega)$ can be attributed solely to
finite-size effects; (ii) the approximation of a single dominating
frequency is a good one for the calculation of the dynamic
susceptibility.

\subsection{Relaxation time}

In Fig.~\ref{9K}(a) we show the frequency dependence of the imaginary
part of the ac susceptibility, measured for a sample of five crystals
of nominally pure CoPhOMe, at a fixed temperature of 9 K and
for $H_{\rm dc}$ ranging from 0.1 to 2
kOe. Fitting of the curves was performed using an extended Debye model
to extract the peak frequency\cite{libromolnano}
\begin{equation}
\label{chis}
\chi(\omega)=\chi_S+{{\chi_T-\chi_S}\over {1+(i\omega \tau})^{1-\alpha}}
\end{equation}
\noindent where $\chi_T$ is the isothermal susceptibility,
$\chi_S$ is the adiabatic susceptibility,
and $\alpha$ is a parameter that measures the distribution of
relaxation times in the sample. In the present measurements it was
found that $\alpha \approx 0$, meaning that the whole system
relaxes with a single characteristic time. The so-obtained field
dependence of the relaxation time, $\tau$, is reported (red
squares) in Fig.~\ref{9K}(b).

The field dependence of the relaxation time at a lower
temperature, $T=5.1$ K, was instead obtained performing dc
measurements. The magnetization was first saturated in $H_{\rm
dc}=25$ kOe, the field was swept to a new $H_{\rm dc}$ value
with a field sweep rate of about 250 Oe/s, and the time dependence of
the magnetization of the system was recorded in presence of an
applied static magnetic field. The relaxation of the magnetization
as a function of time in nonzero field,
measured using a Cryogenic S600
superconducting quantum interference device
(SQUID) magnetometer, is reported in Fig.~\ref{5K}(a). The
magnetization was fitted using a mono-exponential law,
$M(t)=M(t_0)+a_1~e^{-(t-t_0)/\tau}$, and we obtained
the relaxation time $\tau$ versus $H$ reported (red squares) in Figure \ref{5K}(b).

For the interpretation of the $\tau$ versus $H$ data,
concentrations of non-magnetic impurities in the range $c\approx
0.02-0.05$ were assumed, i.e. comparable with those evidenced by
ac and dc susceptibility measurements in the same compound. The
theoretical curves for $\tau$ versus $H$, reported as full
($c=0.02$) and dashed ($c=0.047$) lines in Figure \ref{9K}(b) and
\ref{5K}(b), were determined as follows. The dynamic
susceptibility $\chi(\omega)$ was calculated in the approximation
of a single dominating frequency, see
Eq.~(\ref{chi_omega_single}), and the frequency maximum of
$\chi^{\prime}(\omega)$ was taken to determine
$\tau=1/(\omega_{\rm max})$. In order to reproduce the correct
order of magnitude for the relaxation time $\tau$ of the
magnetization of CoPhOMe (both at $T=5.1$ K and 9 K), one has to
assume a value $\tau_0 \approx 4 \cdot 10^{-13}$ s for the
characteristic time of spin flip of an isolated spin (the only
free parameter in Glauber's theory\cite{Glauber_1963}). For the
sake of simplicity, the same field-independent value was assumed
in all calculations throughout the paper.

As regards the dc field dependence of the relaxation time of the
magnetization, $\tau$, we can conclude that it is well reproduced
by theory, provided that finite-size effects are included. In
fact, in a measurement performed for an inhomogeneous CoPhOMe
sample, no appreciable contribution to the relaxation time,
$\tau$, is expected from the infinite chain at low temperatures,
owing to the high value of the exchange coupling ($J_I/k_B=80$ K).
For example from Eq.~(\ref{lambdainfty}), putting $\tau_0=4 \cdot
10^{-13}$ s, at $T=5$ K one has $\tau_{\infty}=1.25\cdot 10^{15}$
s for $H_{\rm dc}=0$, while in a static field $H_{\rm dc}=10$ Oe
the relaxation time reduces to $\tau_{\infty}=2.77\cdot 10^{-6}$
s. In contrast, an experimentally appreciable relaxation time is
associated with finite-size chains.

\section{Conclusions} \label{Conclusioni}

In this paper, the spin dynamics of the archetypal molecular
magnetic chain, CoPhOMe, in presence of an external magnetic field
of any intensity, was investigated using a simplified model,
consisting of a one-dimensional Ising ferromagnet with a
stochastic dynamics caused by the interaction of the spins with a
heat reservoir.\cite{Glauber_1963}

In the framework of a local-equilibrium approximation, devised to
truncate the infinite sequence of kinetic equations originated by
the presence of a nonzero static magnetic field, and of a linear
response of the system to a small oscillating field, we first
calculated the dynamic susceptibility of an infinite chain.
Next, the theory was generalized to a finite, open Ising chain. We
showed that the dynamic susceptibility of an open chain with a
finite number $N$ of spins can be expressed as a weighted sum of
$N$ frequency contributions, related to the $N$ relaxation rates
of the magnetization fluctuations. From the comparison with the ac
susceptibility data obtained for nominally pure samples we can
draw two conclusions: i) the pure samples are really
non-homogeneous, because regions with very low density of defects
coexist with regions with relevant density of defects; ii) only
the latter regions show a contribution to the dynamic
relaxation, in the frequency range conventionally investigated by
ac susceptibility measurements.

For doped CoPhOMe chains, the approximation of a single dominating
frequency was found to be quite satisfactory in order to account
for ac susceptibility data\cite{Bogani_2005} in a moderate static
magnetic field ($H_{\rm dc}=2$ kOe), both as a function of temperature and of
the frequency of the small oscillating field.

Finally, it is worth observing that, on the basis of our calculation
of the $T$- and $H$-dependence of the relaxation rates (and
corresponding frequency weights), we do not expect that the
approximation of a single dominating frequency will be able
to account for $^1$H nuclear magnetic resonance (NMR)
and muon spin rotation ($\mu$SR) experimental
data\cite{Lascialfari_2003,Micotti_2004,Mariani_2007,Mariani_2008}
in pure and zinc-doped CoPhOMe, because these techniques probe the local
spin dynamics at frequencies ($\nu \approx $ MHz) substantially higher
than the typical frequencies ($\nu \approx $ kHz) used in an
ac susceptibility experiment. Anyway, the present results constitute
a fundamental background for the future theoretical investigation
of such high-frequency regimes.

\begin{acknowledgments}
Fruitful discussions with P. Santini and P. Politi are gratefully
acknowledged. Financial support was provided by MIUR (Italian
Ministry for University and Research) under the 2008 PRIN Program (Contract No. 2008PARRTS 003)
entitled "Topological Effects and Entanglement in Molecular Chains
and Clusters of Spins". L. B. acknowledges
financial support from the German Ministry of Science via DFG, SFB-TRR21
and the Sofja Kovalevskaja award of the Humboldt Stiftung.
\end{acknowledgments}

\begin{widetext}

\appendix

\section{Analytical expressions of the magnetization $M_N$ and
the static susceptibility $\chi_N$ of an open Ising chain of $N$
spins in a dc magnetic field $H_{\rm dc}$}

Following Wortis,\cite{Wortis} the free energy $F_N$ of a finite
chain of $N$ spins, coupled by the Hamiltonian (\ref{IsingFM})
with $H=H_{\rm dc}$ and subject to open boundary conditions, can be
expressed as the sum of a bulk, a surface, and a finite-size
contribution
\begin{equation}
F_N=-k_B T \Bigg\{ N\ln \lambda_0 + \ln\Big({{A_0}\over
{\lambda_0}}\Big) +\ln\Big[ 1+{{A_1}\over
{A_0}}\Big({{\lambda_1}\over {\lambda_0}}\Big)^{N-1}\Big] \Bigg\}
\end{equation}
where
\begin{equation}
\lambda_{0,1}=e^{K}\Big( \cosh h \pm \sqrt{\sinh^2 h +e^{-4K}} \Big)
\end{equation}
\begin{equation}
A_{0,1}=\cosh h \pm {{\sinh^2 h+e^{-2K}}\over {\sqrt{\sinh^2 h+e^{-4K}}}}
\end{equation}
with $K={{J_{\rm I}}\over {k_B T}}$, $h_0={{g \mu_B H_{\rm dc}}\over
{k_B T}}$. For the reader's convenience, explicit expressions for the
static magnetization, $M_N$, and static susceptibility, $\chi_N$,
of a finite open Ising chain with $N$ spins in presence of a
magnetic field are reported hereafter.
The magnetization is
\begin{eqnarray}
\label{MN} M_N &=&-{{\partial F_N}\over {\partial H_{\rm dc}}}=
g\mu_B~ \Bigg\{N {1\over {\lambda_0}}{{\partial
\lambda_0}\over {\partial h_0}} + \Big( {1\over {A_0}}{{\partial
A_0}\over {\partial h_0}} -{1\over {\lambda_0}}{{\partial
\lambda_0}\over {\partial h_0}}\Big) \cr &+&{1\over { \Big[
\big({{\lambda_0}\over {\lambda_1}}\big)^{N-1}+{{A_1}\over {A_0}}
\Big]} } ~\Big[ \Big({1\over {A_0}}{{\partial
A_1}\over {\partial h_0}} -{{A_1}\over {A_0^2}}{{\partial A_0}\over
{\partial h_0}} \Big) +(N-1){{A_1}\over {A_0}} \Big(  {1\over
{\lambda_1}}{{\partial \lambda_1}\over {\partial h_0}} -{1\over
{\lambda_0}}{{\partial \lambda_0}\over {\partial h}} \Big)
\Big]\Bigg\}
\end{eqnarray}
and the static susceptibility is $\chi_N=-\dfrac{\partial^2
F_N}{\partial H_{\rm dc}^2}$
\begin{eqnarray}\label{chiN}
\chi_N&=&-\dfrac{\partial^2
F_N}{\partial H_{\rm dc}^2}=g\mu_B~\Bigg\{N \Big[
-\dfrac{1}{\lambda_0^2}\big(\dfrac{\partial \lambda_0}{\partial
h_0}\big)^2
+\dfrac{1}{\lambda_0}\dfrac{\partial^2\lambda_0}{\partial h_0^2}
\Big]
\cr
&+& \Big[ -\dfrac{1}{A_0^2}\big(\dfrac{\partial A_0}{\partial
h_0}\big)^2 +\dfrac{1}{A_0}\dfrac{\partial^2 A_0}{\partial h_0^2}
+\dfrac{1}{\lambda_0^2}\big(\dfrac{\partial \lambda_0}{\partial
h_0}\big)^2
-\dfrac{1}{\lambda_0}\dfrac{\partial^2\lambda_0}{\partial h_0^2}
\Big] \cr &-&
\dfrac{1}{\Big[(\dfrac{\lambda_0}{\lambda_1})^{N-1}+\dfrac{A_1}{A_0}\Big]^2}
\Big[
(N-1)(\dfrac{\lambda_0}{\lambda_1})^{N-2}\Big(\dfrac{1}{\lambda_1}\dfrac{\partial
\lambda_0}{\partial
h_0}-\dfrac{\lambda_0}{\lambda_1^2}\dfrac{\partial
\lambda_1}{\partial h}\Big)+\dfrac{1}{A_0}\dfrac{\partial
A_1}{\partial h_0}-\dfrac{A_1}{A_0^2}\dfrac{\partial A_1}{\partial
h_0} \Big]\cr &\times& \Big[
 \Big(\dfrac{1}{A_0}\dfrac{\partial
A_1}{\partial h_0}-\dfrac{A_1}{A_0^2}\dfrac{\partial A_0}{\partial
h_0}\Big)+(N-1)\dfrac{A_1}{A_0}\Big(\dfrac{1}{\lambda_1}\dfrac{\partial
\lambda_1}{\partial h_0}-\dfrac{1}{\lambda_0}\dfrac{\partial
\lambda_0}{\partial h_0}\Big) \Big] \cr &+&
\dfrac{1}{\Big[(\dfrac{\lambda_0}{\lambda_1})^{N-1}+\dfrac{A_1}{A_0}\Big]}
\Bigg[\Big(-\dfrac{1}{A_0^2}\dfrac{\partial A_0}{\partial
h_0}\dfrac{\partial A_1} {\partial
h_0}+\dfrac{1}{A_0}\dfrac{\partial^2A_1}{\partial h_0^2}\Big)\cr
&+&\Big(-\dfrac{1}{A_0^2}\dfrac{\partial A_1}{\partial h_0}
\dfrac{\partial A_0}{\partial h_0}+2\dfrac{A_1}{A_0^3}
(\dfrac{\partial A_0}{\partial
h_0})^2-\dfrac{A_1}{A_0^2}\dfrac{\partial^2A_0}{\partial
h_0^2}\Big)\cr &+&(N-1)\Big(\dfrac{1}{A_0}\dfrac{\partial
A_1}{\partial h_0} -\dfrac{A_1}{A_0^2}\dfrac{\partial A_0}{\partial
h_0}\Big) \Big(\dfrac{1}{\lambda_1}\dfrac{\partial
\lambda_1}{\partial h_0} -\dfrac{1}{\lambda_0}\dfrac{\partial
\lambda_0}{\partial h_0}\Big)\cr &+&(N-1)\dfrac{A_1}{A_0}
\Big(-\dfrac{1}{\lambda_1^2}(\dfrac{\partial \lambda_1}{\partial
h_0})^2+\dfrac{1}{\lambda_1}\dfrac{\partial^2\lambda_1}{\partial
h_0^2}+\dfrac{1}{\lambda_0^2}(\dfrac{\partial \lambda_0}{\partial
h_0})^2-\dfrac{1}{\lambda_0}\dfrac{\partial^2\lambda_0}{\partial
h_0^2}\Big) \Bigg] \Bigg\}
\end{eqnarray}

The derivatives with respect to the magnetic field in
Eq.~(\ref{MN}) and Eq.~(\ref{chiN}) are expressed as
\begin{equation}
{{k_B T}\over {g\mu_B} } {{\partial \lambda_{0,1}}\over {\partial
H_{\rm dc}}}=e^K\Big( \sinh
h \pm {{\sinh (2h_0)}\over {2\sqrt{\sinh^2 h_0 +e^{-4K}}}} \Big)
\end{equation}
\begin{equation}
{{k_B T}\over {g \mu_B}}{{\partial A_{0,1}}\over {\partial
H_{\rm dc}}}=\sinh h
 \pm {{\sinh (2h_0)}\over {\sqrt{\sinh^2 h_0 +e^{-4K}}}}
 \Big[ 1-  { {\sinh^2 h_0+e^{-2K} }\over { 2\big( \sinh^2 h_0 +e^{-4K}\big) }}\Big]
\end{equation}

\begin{eqnarray}
\Big({{k_B T}\over {g\mu_B}}\Big)^2\dfrac{\partial^2
\lambda_{0,1}}{\partial H_{\rm dc}^2}&=&e^{K}\cosh h_0 \cr &\pm& \Biggr[\dfrac{ e^{2K}\cosh
(2h_0)}{\sqrt{e^{2K}\sinh^2 h_0 +e^{-2K}}}- \dfrac{1}{4}\dfrac{
e^{4K}\sinh^2 (2h_0)}{\Big(e^{2K}\sinh^2 h_0 +e^{-2K}\Big)^{3/2}}
\Biggr]
\end{eqnarray}

\begin{eqnarray}
&&\Big({{k_B T}\over {g\mu_B}}\Big)^2\dfrac{\partial^2
A_{0,1}}{\partial H_{\rm dc}^2}=\cosh h_0 \pm \Bigg\{\dfrac{2\cosh (2h_0)}{\sqrt{\sinh^2
h+e^{-4K}}} -\dfrac{ \sinh^2 (2h_0)}{\Big(\sinh^2 h_0
+e^{-4K}\Big)^{3/2}} \cr &+&\dfrac{3}{4}\dfrac{\sinh^2 h_0+e^{-2K}}{\big(\sinh^2
h_0+e^{-4K}\big)^{5/2}} \sinh^2 (2h_0)-\dfrac{\sinh^2
h_0+e^{-2K}}{\big(\sinh^2 h_0+e^{-4K}\big)^{3/2}} \cosh
(2h_0)\Biggr]\Bigg\}
\end{eqnarray}

\section{Relaxation times of an open chain of $N$ Ising spins in
an applied magnetic field}

In order to calculate the relaxation times of an open chain of $N$
spins coupled by the Ising exchange Hamiltonian and subject to a
static magnetic field $H$, Eq.~(\ref{IsingFM}), following
Glauber\cite{Glauber_1963} one starts by writing the kinetic
equation of motion for the time-dependent average of an interior
spin ($\langle \sigma_p \rangle$ with $1<p<N$)
\begin{equation}
\label{bulk} \tau_0 {{d\langle \sigma_p \rangle}\over
{dt}}=-\langle \sigma_p \rangle+{{\gamma}\over 2} \Big( \langle
\sigma_{p-1}\rangle + \langle \sigma_{p+1}\rangle \Big)+\tanh h(t)
\Big[ 1-{{\gamma}\over 2} \Big( \langle \sigma_{p-1} \sigma_p
\rangle + \langle \sigma_p \sigma_{p+1} \rangle \Big)\Big]
\end{equation}
where $\tau_0$ is the characteristic time for the spin flip of an
isolated spin, $\gamma=\tanh(2K)$ and $h(t)={{g \mu_B H(t)}\over
{k_B T}}$. For the two spins ($p=1$ and $p=N$) located at the ends
of the open chain, following Coulon {\it et al.}\cite{Coulon}
one has instead
\begin{equation}
\label{end1} \tau_0 {{d\langle \sigma_1 \rangle}\over
{dt}}=-\langle \sigma_1 \rangle+\eta \langle \sigma_2 \rangle
+\tanh h(t) \Big[1- \eta \langle \sigma_1 \sigma_2 \rangle \Big]
\end{equation}
\begin{equation}
\label{endN} \tau_0 {{d\langle \sigma_N \rangle}\over
{dt}}=-\langle \sigma_N \rangle+\eta \langle \sigma_{N-1} \rangle
+\tanh h(t) \Big[1- \eta \langle \sigma_{N-1} \sigma_N \rangle
\Big]
\end{equation}
where $\eta=\tanh K$.\cite{Glauber_1963} In our case,
see Eq.~(\ref{HDCAC}), the time-dependent magnetic field is
assumed to be $H(t)=H_{\rm dc}+H_1~ e^{-i\omega t}$ i.e. the sum of a
static dc field of any intensity, $H_{\rm dc}$, and of a much lower AC
field, $H_1 \ll H_{\rm dc}$, oscillating at the angular frequency
$\omega$. Thus we can expand $\tanh h(t) \approx \tanh h_0 + h_1
e^{-i \omega t}(1-\tanh^2 h_0)$, where $h_0={{g \mu_B H_{\rm dc}}\over
{k_B T}}$ and $h_1={{g \mu_B H_1}\over {k_B T}}$.

Following Coulon {\it et al.},\cite{Coulon} a linearization of the
equations (\ref{bulk}), (\ref{end1}), and (\ref{endN}) around the
spin equilibrium values is next performed in the framework of
linear response theory (i.e. small departures from thermal
equilibrium are assumed)
\begin{eqnarray}
\langle \sigma_p \rangle &\approx& \langle
\sigma_p\rangle_{N,eq}+\delta \langle \sigma_p \rangle \cr \langle
\sigma_p \sigma_{p+1} \rangle &\approx& \langle \sigma_p
\sigma_{p+1} \rangle_{N,eq}+\delta \langle \sigma_p
\sigma_{p+1}\rangle
\end{eqnarray}
In this way, we obtain a system of $N$ differential equations in
the spin fluctuations ($\delta \langle \sigma_p\rangle$ with
$p=1,\cdots,N$) where, on the right hand sides, the presence of
the field involves the presence of variations of time-dependent
nearest neighbors spin-spin correlation functions ($\delta \langle
\sigma_p \sigma_{p+1} \rangle$)
\begin{eqnarray}
\tau_0 {{d \delta \langle \sigma_1 \rangle}\over
{dt}}&=&-\delta\langle \sigma_1 \rangle+\eta \Big[ \delta\langle
\sigma_2 \rangle - \tanh h_0 ~\delta \langle \sigma_1 \sigma_2
\rangle \Big] +h_1 e^{-i\omega t}(1-\tanh^2 h_0) \Big[1- \eta
\langle \sigma_1 \sigma_2 \rangle_{N,eq} \Big] \cr \tau_0
{{d\delta\langle \sigma_p \rangle}\over {dt}}&=& -\delta\langle
\sigma_p \rangle +{{\gamma}\over 2} \Big[ \Big( \delta\langle
\sigma_{p-1}\rangle +\delta\langle \sigma_{p+1}\rangle \Big) -
\tanh h_0 \Big(\delta \langle \sigma_{p-1}\sigma_p\rangle
+\delta\langle\sigma_p \sigma_{p+1}\rangle \Big)\Big]\cr &+&h_1
e^{-i\omega t} (1-\tanh^2 h_0) \Big[ 1-{{\gamma}\over 2} \Big(
\langle \sigma_{p-1} \sigma_p \rangle_{N,eq} + \langle \sigma_p
\sigma_{p+1} \rangle_{eq} \Big)\Big] \cr \tau_0 {{d\delta\langle
\sigma_N \rangle}\over {dt}}&=&-\delta\langle \sigma_N
\rangle+\eta \Big[ \delta\langle \sigma_{N-1} \rangle -\tanh h_0~
\delta\langle \sigma_{N-1}\sigma_N\rangle \Big]+h_1 e^{-i\omega
t}(1-\tanh^2 h_0) \Big[1- \eta \langle \sigma_{N-1} \sigma_N
\rangle_{N,eq} \Big]
\end{eqnarray}
If one writes down the kinetic equations for $\delta \langle
\sigma_p \sigma_{p+1} \rangle$, one finds and infinite sequence of
equations involving other, higher-order spin correlation
functions. Such an infinite hierarchy of equations can be
decoupled by resorting to the local-equilibrium approximation
first proposed by Huang in the case of the kinetic equation of an
infinite Ising chain in a magnetic field:\cite{Huang} i.e. the
relation existing between the magnetization and the correlation
function at thermal equilibrium\cite{Marsh} is assumed to hold
{\it locally} although the system is not in equilibrium ($t\ne
0$). The advantage of the local-equilibrium approximation with
respect to perturbative methods\cite{Glauber_1963} or mean field
approximation is that it provides an exact steady-state
solution.\cite{Huang}

In the case of a finite chain with $N$ spins, the
local-equilibrium approximation was implemented, following Coulon
et al.,\cite{Coulon} by expressing the variations of two-spin
correlation functions in terms of a linear combination of the
variations of single-spin averages
\begin{equation}
\delta \langle \sigma_p \sigma_{p+1} \rangle = A_{N,p} \delta
\langle \sigma_p \rangle +B_{N,p}\delta \langle \sigma_{p+1}
\rangle.
\end{equation}
The coefficients are\cite{Coulon}
\begin{equation}
A_{N,p}={{\langle \sigma_{N-p} \rangle_{N-p,eq} -\eta
~\langle\sigma_{p}\rangle_{p,eq}}\over
{1-\eta~\langle\sigma_p\rangle_{p,eq}\langle\sigma_{N-p}\rangle_{N-p,eq}}}
,~~~~~ B_{n,p}={{\langle \sigma_{p} \rangle_{p,eq} -\eta~
\langle\sigma_{N-p}\rangle_{N-p,eq}}\over
{1-\eta~\langle\sigma_p\rangle_{p,eq}\langle\sigma_{N-p}\rangle_{N-p,eq}}}
\end{equation}
where the average spin values can be calculated, at thermal
equilibrium, by using the recursive relation\cite{Matsubara}
\begin{equation}
\langle \sigma_p \rangle_{p,eq} ={{\tanh h_0+\eta~ \langle
\sigma_{p-1}\rangle_{p-1,eq}}\over {1+ \eta ~\tanh h_0~\langle
\sigma_{p-1} \rangle_{p-1,eq}}}
\end{equation}
with the initial condition $\langle \sigma_1 \rangle_{1,eq}= \tanh
h_0$. The set of $N$ linear differential equations in the $N$
variables $\delta \langle \sigma_p \rangle$ can be rewritten in
matrix form as
\begin{equation}
\label{eqmatrix} \tau_0 {{d{\bf \Sigma}}\over {dt}}=-{\bf Y}\cdot {\bf \Sigma}+h_1
e^{-i\omega t}(1-\tanh^2 h_0) {\bf \Psi}.
\end{equation}
The $N\times 1$ vector ${\bf \Sigma}$ has elements $\Sigma_p=\delta
\langle \sigma_p \rangle$ ($p=1,...,N$). ${\bf Y}$ is a real, symmetric,
tridiagonal, $N\times N$  matrix whose nonzero elements are
($1<p<N$)
\begin{eqnarray}
\label{matrix_elements_Y}
Y_{1,1}=1+\eta~A_{N,1}~\tanh h_0 ,~Y_{1,2}&=&\eta~ \Big( B_{N,1}~\tanh h_0
 -1 \Big),\cr Y_{p-1,p}={{\gamma}\over 2} \Big(A_{N,p-1}~\tanh h_0
-1\Big),~Y_{p,p}&=&1+{{\gamma}\over 2} ~ \Big(
A_{N,p}+B_{N,p-1}\Big)~\tanh h_0,~Y_{p,p+1}={{\gamma}\over 2}\Big(B_{n,p}~\tanh
h_0-1 \Big),\cr Y_{N,N-1}&=&\eta~\Big(A_{N,N-1}
~\tanh h_0 -1\Big),~Y_{N,N}=1+\eta~B_{N,N-1}~ \tanh h_0
\end{eqnarray}
The $N\times 1$ vector ${\bf \Psi}$ in Eq.~(\ref{eqmatrix}) has elements
\begin{eqnarray}
\label{inhomogeneous} \Psi_1&=&1-\eta ~\langle \sigma_1\sigma_2
\rangle_{N,eq}\cr \Psi_p&=&1-{{\gamma}\over 2}~ \Big[\langle
\sigma_{p-1}\sigma_p\rangle_{N,eq}+
\langle\sigma_p\sigma_{p+1}\rangle_{N,eq}\Big],~~~~~1<p<N\cr
\Psi_N&=&1-\eta ~\langle\sigma_{N-1}\sigma_N\rangle_{N,eq}=\Psi_1.
\end{eqnarray}
This equation can be solved, e.g. using the method of
eigenfunctions,\cite{vanKampen} as described in Section IV.B.

\end{widetext}

\end{document}